%% file: main.tex
\documentclass[prd, aps, floatfix, 10pt, twocolumn, superscriptaddress, preprintnumbers,nofootinbib]{revtex4-2}

\input{packages}

\input{defs}

\begin{document}

\title{Saturation effects in diffractive deep inelastic scattering}

\author{Jani Penttala}
\email{janipenttala@physics.ucla.edu}
\affiliation{
Department of Physics and Astronomy, University of California, Los Angeles, CA 90095, USA
}
\affiliation{Mani L. Bhaumik Institute for Theoretical Physics, University of California, Los Angeles, California 90095, USA}

\author{Christophe Royon}
\email{christophe.royon@cern.ch
}
\affiliation{Department of Physics and Astronomy, The University of Kansas, Lawrence KS 66045, USA
}

\author{Jarno Vierros}
\email{jarno.johannes.vierros@cern.ch
}
\affiliation{Department of Physics and Astronomy, The University of Kansas, Lawrence KS 66045, USA}
\affiliation{Departement of Physics, The University of Helsinki, Helsinki, Finland}
%\todo{}

\begin{abstract}
We study the production of $c$ and $b$ quarks in photon-induced processes as a probe of gluon saturation, with a focus on the high-energy region accessible in ultra-peripheral collisions (UPCs) at the LHC. We compare predictions for inclusive and diffractive heavy quark production with and without saturation effects by using the nonlinear Balitsky--Kovchegov and the linear Balitsky--Fadin--Kuraev--Lipatov equations for the evolution of the dipole amplitude. The percentage of coherent diffractive events is predicted to be $~20\%$ in the saturation model, which is approximately twice as large as predictions based on parton distribution functions and shadowing. This indicates that measurements of diffractive heavy quark production in UPCs may provide a sensitive probe for distinguishing linear and nonlinear regimes of quantum chromodynamics.
\end{abstract}

\maketitle

\section{Introduction}
One of the key discoveries of the H1 and ZEUS experiments at HERA was the rapid increase of the gluon density at small Bjorken-$x$, where $x$ represents the fraction of the proton momentum carried by the struck quark~\cite{H1:2015ubc}. At very small values of $x$, this rapid increase of the gluon density
would eventually violate unitarity, indicating that nonlinear QCD effects become important and lead to gluon saturation~\cite{Gelis:2010nm}.
However, while this prediction from QCD is theoretically well understood, it has eluded definite experimental verification due to the smallness of the saturation effects at the currently accessible energies~\cite{Morreale:2021pnn}.
The search for gluon saturation serves as the key motivation for many current data analyses done at the LHC, and it is one of the major science goals of the future Electron-Ion Collider~\cite{Accardi:2012qut,AbdulKhalek:2021gbh}.

The interaction with the highly gluonic target can be described within the framework of color glass condensate (CGC), where the interaction is described by non-perturbative Wilson-line correlators. 
For the two-point Wilson-line correlator, known as the dipole amplitude, the evolution is given by the Balitsky--Kovchegov (BK) equation~\cite{Balitsky:1995ub,Kovchegov:1999yj}, which includes saturation effects. 
To estimate the importance of saturation, this can be compared to the linear
Balitsky--Fadin--Kuraev--Lipatov (BFKL) evolution~\cite{Lipatov:1976zz,Kuraev:1977fs,Balitsky:1978ic}
where the nonlinear saturation effects are not included.
The differences between these two evolution equations can then be directly attributed to gluon saturation, making it possible to study the onset of gluon saturation at different energies and momentum scales of the process.

To probe saturation effects, it  is necessary to consider a process where the typical momentum scale is large enough to be perturbative but small enough to be comparable to the saturation scale $Q_s$.
 At HERA, $Q_s$ was estimated to be smaller than $\SI{1}{GeV}$, which is too low for perturbative QCD to be valid in the saturation region~\cite{H1:2015ubc,Quiroga-Arias:2012uul}.
It is, however, possible to be sensitive to much higher values of $Q_s$ at the LHC
due to the higher energies achieved at ultra-peripheral collision (UPCs) and due to the nuclear enhancement of the saturation scale, $Q_s^2 \sim A^{1/3}$, where $A$ is the mass number of the nucleus.
This makes UPC processes a natural candidate in the search for saturation.

In our previous work~\cite{Penttala:2024hvp}, we presented a comparison between recent measurements of exclusive $J/\psi$ and $\Upsilon$ production in UPCs with predictions using the BFKL and the BK evolution equations. 
Because of the lower mass of the $J/\psi$ particle, the $J/\psi$ production data with nuclear targets clearly favor models with saturation. However, alternative models based on nuclear parton distribution functions (PDFs) and nuclear shadowing can also describe the observed data~\cite{Paakkinen:2026dnh,Guzey:2025jfj}. 
The goal of this paper is thus to find an observable that could discriminate between both approaches. We propose to measure the production of $c$ and $b$ quarks in ultra-peripheral $p+\text{Pb}$ and $\text{Pb}+\text{Pb}$ collisions, and especially in coherent diffractive events where the target nucleon or nucleus is intact after the collision. Nuclear PDF and shadowing models predict a percentage of diffractive events for $c \bar{c}$ and $b \bar{b}$ production to be similar to what was measured at HERA, about 12\%\cite{H1:2012pbl,H1:2006zxb,Paakkinen:2026dnh,Guzey:2025jfj}. 
As we will show, saturation based models show a much higher fraction of diffractive events at high energies.

In Sec.~\ref{sec:theory}, we give the theoretical framework of our calculation, and in Sec.~\ref{sec:numerics} the numerical results including the comparison with measurements at HERA and the predictions for the LHC energies. 
We present a discussion of the results and conclude in Sec.~\ref{sec:discussion}.

\section{Theoretical framework}
\label{sec:theory}

\begin{figure*}[t]
	\centering
    \begin{subfigure}[T]{0.45\textwidth}
        \centering
        \includegraphics[width=\textwidth]{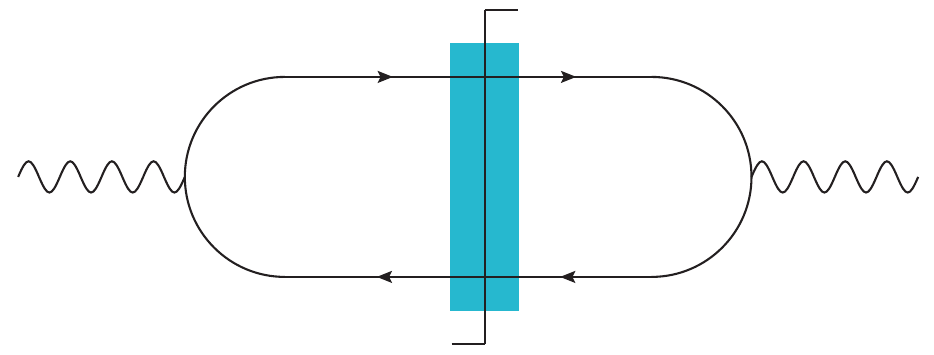}
        \caption{ Inclusive DIS. }
        \label{fig:inclusive_dis}
    \end{subfigure}
    \begin{subfigure}[T]{0.45\textwidth}
        \centering
        \includegraphics[width=\textwidth]{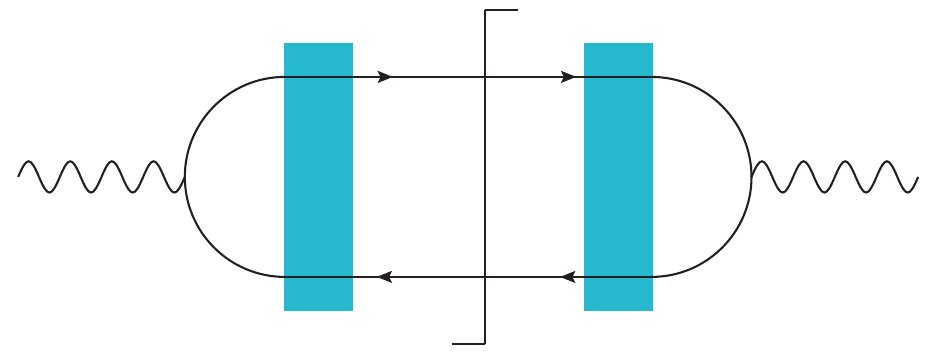}
        \caption{ Diffractive DIS. }
        \label{fig:diffractive_dis}
    \end{subfigure}
    \caption{
    Relevant Feynman diagrams for inclusive and diffractive DIS in the CGC framework.
    The blue rectangle corresponds to the eikonal scattering off the target nucleon or nucleus, and the cut represents the final state.
    }
    \label{fig:cgc_diagrams}
\end{figure*}

\subsection{Dipole picture}

Deep inelastic scattering at small $x$ can be described using the dipole picture, which factorizes the non-perturbative scattering off the target from the rest of the process.
At leading order, the incoming virtual photon first fluctuates into a quark--antiquark dipole which interacts with the target proton or nucleus eikonally, see Fig.~\ref{fig:inclusive_dis}.
The formation of the quark--antiquark dipole can be calculated perturbatively, which leads to the following production cross section into a specific quark flavor $f$:
\begin{equation}\label{eq:inclusive_quark_production}
\begin{split}
     &\sigma^{\gamma_\lambda^* + p/A \to q_f \bar q_f + X}
     \\
     =&\frac{4 \aem \nc e_f^2}{(2\pi)^2} \int \dd[2]{\rt}\dd[2]{\bt}
     \int_0^1 \dd{z} \mathcal{K}_\lambda(\rt,z)
     N_x(\xt,\yt),
\end{split}
\end{equation}
where $\rt$ is the transverse separation of the $q \bar q$ dipole, $\bt$ is the impact parameter, and $z$ is the fraction of the photon's longitudinal momentum carried by the quark. 
The transverse coordinates $\xt$ and $\yt$ of the quark and antiquark can be related to the transverse separation $\rt$ and the impact parameter $\bt$ as
\begin{align}
    \rt &= \xt - \yt,
    &
    \bt &= z \xt + (1-z) \yt.
\end{align}
In Eq.~\eqref{eq:inclusive_quark_production}, $\lambda$ denotes the polarization of the initial photon, with the functions $\mathcal{K}_\lambda$ given as
\begin{equation}
\begin{split}
    &\mathcal{K}_L(\rt, z) = 4 Q^2 z^2 (1-z)^2 K_0(\epsilon \abs{\rt})^2,\\
    &\mathcal{K}_T(\rt, z) = m_f^2 K_0(\epsilon \abs{\rt})^2 + \epsilon^2 \qty[ z^2 + (1-z)^2]  K_1(\epsilon \abs{\rt})^2
\end{split}
\end{equation}
for longitudinally and transversely polarized photons.
We have also denoted $\epsilon^2 = m_f^2 + z (1-z)Q^2$, with $m_f$ being the quark mass 
and $Q^2$ the virtuality of the incoming photon.
In the numerical implementation, the mass of the light quarks is taken to zero, $m_l = \SI{0}{GeV}$ ($l=u,d,s$),
and the charm and bottom quark masses are given as 
$m_c = \SI{1.27}{GeV}$ and $m_b = \SI{4.18}{GeV}$, respectively.
The inclusive $\gamma + p/A $ cross section can then be calculated by summing over the quark flavors.

The dipole amplitude $N$, describing the scattering of the $q \bar q$ dipole off the target, is defined as
\begin{equation}
    N_x(\xt ,\yt) = 1 - \frac{1}{\nc} \Tr\expval{ V(\xt) V^\dag(\yt) }_x,
\end{equation}
where $V$ and $V^\dag$ correspond to Wilson lines describing the scattering of the quark and antiquark off the target,
and $\expval{\ldots}$ is the CGC average over the target color configurations.
The $x$-variable in the dipole amplitude is given by
\begin{equation}
\label{eq:xm}
    x = \frac{Q^2 + 4 m_f^2}{W^2 + Q^2 },
\end{equation}
where $W^2$ is the center-of-mass energy of the photon--target system.
Note that for massless quarks this agrees with the standard Bjorken-$x$ variable.

For diffractive DIS, we also require a rapidity gap in the final state.
This kind of scattering can be theoretically understood as a color-neutral exchange between the $q \bar q$ dipole and the target, which can be imposed by demanding that the outgoing diffractive state is a color singlet.
The corresponding Feynman diagram is shown in Fig.~\ref{fig:diffractive_dis}, which results in the following cross section
\begin{equation}\label{eq:differential_diffractive_quark_production}
\begin{split}
    &\frac{\dd{}}{\dd{M_X^2}} \sigma^{\gamma_\lambda^* + p/A \to q_f \bar q_f + X}_\text{D}
    \\
    =& \frac{1}{(2\pi)^3}\alpha_\text{em} N_c 
    e_f^2
    \int_0^1 \dd{z} 
    \int \dd[2]{\rt} \dd[2]{\ov \rt}
     \dd[2]{\bt} \dd[2]{\ov\bt} \delta^{(2)}(\bt -\ov \bt)
    \\
    &\times
    z(1-z)  J_0\qty( \sqrt{z (1-z)M_X^2 - m_f^2}\abs{ \rt -\ov \rt } )
    \\
    &\times
     \theta\qty(z (1-z)M_X^2 - m_f^2)
    \mathcal{H}_\lambda(\rt, \ov \rt , z)
    N_{\xpom}(\xt,\yt)
    N_{\xpom}(\ov \xt,\ov \yt),
\end{split}
\end{equation}
measured differentially in terms of the invariant mass $M_X^2$ of the diffractive final state.
The coordinates $\ov \xt$ and $\ov \yt$ correspond to the quark and antiquark in the conjugate amplitude, and they are related to the integration variables $\ov \rt$ and $\ov \bt$ with
\begin{align}
    \ov \rt &= \ov \xt - \ov \yt,
    &
    \ov \bt &= z \ov \xt + (1-z) \ov \yt.
\end{align}
The impact factors $\mathcal{H}_\lambda$ for different incoming photon polarizations are given as:
\begin{equation}
\begin{split}
    \mathcal{H}_L(\rt, \ov \rt, z) =& 4 Q^2 z^2 (1-z)^2 K_0(\epsilon \abs{\rt}) K_0(\epsilon \abs{\ov \rt}),\\
    \mathcal{H}_T(\rt, \ov \rt, z) =& m_f^2 K_0(\epsilon \abs{\rt}) K_0(\epsilon \abs{\ov \rt})
    \\
    &
    + \epsilon^2 \qty[ z^2 + (1-z)^2] \frac{\rt \vdot \ov \rt}{\abs{\rt}\abs{\ov \rt}} K_1(\epsilon \abs{\rt}) K_1(\epsilon \abs{\ov \rt}).
\end{split}
\end{equation}
For diffraction, the relevant $x$-variable for the dipole amplitude is given as
\begin{equation}
    \xpom = \frac{Q^2 + M_X^2 + 4m_f^2}{W^2 + Q^2},
\end{equation}
and it is also customary to define the variable
\begin{equation}
    \beta = \frac{Q^2}{Q^2+M_X^2+ 4m_f^2}.
\end{equation}
The variables $\xpom$ and $\beta$ have a physical meaning in the parton distribution approach to diffraction, where $\xpom$ describes the momentum of the target carried by the pomeron, and $\beta$ gives the momentum fraction of the pomeron carried by the scattering parton.

Finally, we note that the diffractive cross section can also be integated over the invariant mass of the final state $M_X^2$, giving us
\begin{equation}\label{eq:diffractive_quark_production}
\begin{split}
    &\sigma^{\gamma_\lambda^* + p/A \to q_f \bar q_f + X}_\text{D}
    \\
    =&\frac{2 \aem \nc e_f^2 }{\left(2\pi\right)^2}
    \int_0^1 \dd{z}
    \int \dd[2]{\rt} \int \dd[2]{\bt}
   \mathcal{K}_\lambda(\rt, z)N_x^2\qty(\rt, \bt ),
\end{split}
\end{equation}
where the dipole amplitude now depends on $x$ defined in Eq.~\eqref{eq:xm}\footnote{While the difference on using $x$ or $\xpom$ is formally of higher order in perturbation theory, it is expected that using $x$ leads to a better convergence of the perturbative series after the invariant mass $M_X$ has been integrated out.}.
Notably, the difference between inclusive and diffractive DIS corresponds to changing $2 N \mapsto N^2$ for the interaction with the target~\cite{Kovchegov:2012mbw}.
This also means that the maximum amount of diffractive events is achieved in the unitarity limit $N = 1$, where
$\sigma_D / \sigma = 1/2$.

Instead of measuring cross sections for different photon polarizations, the experimental data is usually defined in terms of structure functions.
For inclusive DIS, these are given by
\begin{align}
    F_\lambda(x,Q^2) &= \frac{Q^2}{4\pi^2 \aem} \sigma^{\gamma^*_\lambda +p/A}, \\
        F_2(x,Q^2) &= F_L(x,Q^2) + F_T(x,Q^2).
\end{align}
The data is then usually given in terms of the reduced cross section, defined as:
\begin{equation}
    \sigma_\text{red}(y,x,Q^2) = F_2(x,Q^2) - \frac{y^2}{1+(1-y)^2} F_L(x,Q^2),
\end{equation}
where $y$ is the inelasticity parameter.
For diffraction, the related structure functions are defined as:
\begin{align}
  \xpom  F^{\text{D}(3)}_\lambda(x,Q^2, \beta) 
  =&
  \frac{Q^2}{4\pi^2 \aem} 
  \frac{Q^2}{\beta}
  \frac{\dd{\sigma_\text{D}^{\gamma^*_\lambda +p/A}}}{\dd{M_X^2}}, 
  \\
  \begin{split}
   \xpom    F^{\text{D}(3)}_2(x,Q^2,\beta) =& 
   \xpom    F^{\text{D}(3)}_L(x,Q^2,\beta) \\
   &+ \xpom    F^{\text{D}(3)}_T(x,Q^2,\beta),    
   \end{split}
\end{align}
where we have now introduced the additional parameter $\beta$.

\subsection{Dipole amplitude}

The dipole amplitude describing the interaction with a gluonic target is a non-perturbative quantity that has to be modeled.
However, its energy dependence is perturbative and governed by the BK equation~\cite{Balitsky:1995ub,Kovchegov:1999yj} in the large-$\nc$ limit:
\begin{equation}
\begin{split}
        &\frac{\partial}{\partial \log 1/x} N_x(\xt,\yt)
        \\
        =&  \int \dd[2]{\zt} \mathcal{K}(\xt,\yt,\zt) 
        \bigl[  
        N_x\qty(\xt,\zt ) + N_x(\zt,\yt)
        - N_x(\xt,\yt) 
        \\
        &
        - \delta_\text{sat} N_x\qty(\xt,\zt )  N_x(\zt,\yt)
        \bigr]
\end{split}
\end{equation}
with $\delta_\text{sat} = 1$.
The BK evolution incorporates the gluon saturation phenomenon in the nonlinear term $ N_x\qty(\xt,\zt ) N_x(\zt,\yt)$.
This can be compared to the linear BFKL~\cite{Lipatov:1976zz,Kuraev:1977fs,Balitsky:1978ic} evolution, where $\delta_\text{sat} = 0$, such that the nonlinear term is absent.
Deviations between BK and BFKL evolutions can then be directly attributed to gluon saturation.
The kernel $\mathcal{K}$ is given as
\begin{equation}
\begin{split}
\mathcal{K}(\xt_0,\xt_1,\xt_2)=&
    \frac{N_c}{2\pi^2} \Bigl[ \sqrt{\as(\xt_{20})} m'\abs{\xt_{20}} K_1(m'|\xt_{20}|) 
    \frac{\xt_{20}}{\xt_{20}^2}
    \\
    &
    -\sqrt{\as(\xt_{21})} m'\abs{\xt_{21}} K_1(m'|\xt_{21}|) \frac{\xt_{21}}{\xt_{21}^2}
    \Bigr]^2,    
\end{split}
\end{equation}
where we have used the compact notation $\xt_{ij} \equiv \xt_i - \xt_j$.
This corresponds to the daughter dipole description for the running of the coupling~\cite{Lappi:2012vw,Altinoluk:2023krt,Kovner:2023vsy}, and we have also included an infrared regulator $m' = \SI{0.4}{GeV}$~\cite{Schlichting:2014ipa,Mantysaari:2018zdd,Mantysaari:2024zxq} to suppress the non-perturbative contributions in the evolution that would otherwise lead to non-physical Coulomb tails~\cite{Kovner:2001bh,Kovner:2002xa,Kovner:2002yt}. 
The coordinate-space running coupling in the evolution is given by~\cite{Mantysaari:2024zxq}
\begin{equation}
    \alpha_s(r^2)
    = \frac{4 \pi}{\beta_0 
    \log \qty[   \qty(\frac{\mu_0^2}{\lambdaQCD^2})^{1 / \zeta}+
     \qty(\frac{4}{r^2\lambdaQCD^2})^{1 / \zeta}]^\zeta
    },
\end{equation}
where $\beta_0 = (11N_c - 2 N_f)/3 $,
the active number of flavors is chosen as
$N_f = 3$,
$\lambdaQCD = \SI{0.025}{GeV}$ to match the energy dependence of the exclusive $J/\psi$ production data~\cite{Mantysaari:2024zxq},
and the parameters
$\mu_0 = \SI{0.28}{GeV}$ and $\zeta= 0.2$
regulate the infrared behavior of the coupling constant~\cite{Lappi:2012vw,Mantysaari:2018zdd,Mantysaari:2022sux,Mantysaari:2024zxq}.

For the non-perturbative initial condition,
we use the same model for both BK and BFKL evolutions to study the saturation effects arising from the evolution itself.
Following Refs.~\cite{Mantysaari:2024zxq,Penttala:2024hvp}, we use an impact-parameter-dependent initial condition defined as:
\begin{equation}
\label{eq:N_ip}
N(\xt, \yt) 
=1 -  \exp(
    - \kappa \int \dd[2]{\zt}  T(\zt)  \Gamma_{\zt}(\xt,\yt)  ) \,,
\end{equation}
where 
\begin{equation}
\label{eq:Gamma}
\Gamma_{\zt}(\xt,\yt) = 
    \bigl[ K_0( m \abs{\xt-\zt} )  -  K_0( m \abs{\yt-\zt} ) \bigr]^2 \,,
\end{equation}
$m = \SI{0.4}{GeV}$ is an infrared regulator and $\kappa = 0.66$ describes the strength of the interaction with the target.
The shape of the target is given by the thickness function $T(\zt)$,
which for protons we model as a Gaussian:
\begin{equation}
\label{eq:gaussian}
    T_p(\bt) = \frac{1}{2\pi B_p} e^{-\bt^2/(2B_p)} \,.
\end{equation}
with $B_p = \SI{3}{GeV^{-2}}$.
These parameter values are found to be in good agreement with exclusive $J/\psi$ production data for proton  targets~\cite{Mantysaari:2024zxq}.
For nuclear targets, we follow the optical Glauber approach and model it as a collection of nucleons.
As a result, we only need to change
the thickness function to a Woods--Saxon distribution according to
\begin{equation}
 T_p(\bt) \mapsto    AT_A(\bt) =A \int \dd{z} \rho_A(\bt, z) \,,
\end{equation}
where
\begin{equation}
    \rho_A(\bt, z)
    = \frac{n}{ 1+ \exp[ \frac{\sqrt{\bt^2 + z^2 } - R_A}{d} ]} \,.
\end{equation}
Here, $n$ is a normalization constant such that $\int \dd[2]{\bt} \dd{z} \rho_A(\bt,z) =1$, and the parameters describing the nuclear shape are taken as $d = \SI{0.54}\,\si{fm}$ and $R_A = ( 1.12 A^{1/3} - 0.86 A^{-1/3} )\,\si{fm}$~\cite{Lappi:2013zma}.

For predictions of diffractive DIS, we use the same dipole amplitude as in Refs.~\cite{Mantysaari:2024zxq,Penttala:2024hvp}.
For inclusive DIS, the parameter $\kappa$ is changed to $\kappa = 0.25$ to agree with the overall normalization of the experimental data.
While both inclusive and diffractive DIS should be described by the same dipole amplitude, our purpose here is to compare saturation effects when going from protons to nuclear targets, such that the parameters are consistent within each observable.
Similar problems in the simultaneous description of both inclusive DIS and exclusive $J/\psi$ production have also been noted in Ref.~\cite{Mantysaari:2018zdd}, where the difference was attributed to the different behavior at large dipole sizes that is sensitive to the description of non-perturbative physics.
The discrepancy between inclusive and diffractive DIS could also be due to higher-order corrections in perturbation theory, which can generally be large at small $x$~\cite{Mantysaari:2021ryb,Mantysaari:2022bsp,Mantysaari:2022kdm,Mantysaari:2024zxq,Ducloue:2017ftk,Kaushik:2025roa}.
Another possibility is considering a different model for the dipole amplitude, especially a different proton shape $T_p(\bt)$.
In general, the proton shape can affect the overall normalization between diffractive and inclusive events, as can be seen in the dilute approximation where the inclusive cross section scales as $\sim \int \dd[2]{\bt} T_p(\bt)$ while the diffractive cross section scales as $\sim \int \dd[2]{\bt} T_p(\bt)^2$~\cite{Lappi:2023frf}.
However, we have checked that considering a more general parametrization for the shape as in Ref.~\cite{Lappi:2023frf} does not affect the overall normalization once it has been fitted to the $t$-differential exclusive $J/\psi$ production data.
We also note that in the past, the so-called real-part and skewness corrections have been applied to diffractive processes~\cite{Kowalski:2006hc}, but leave them out in this work due to their dubious nature in the small-$x$ formalism~\cite{Kovchegov:2026gwb}.
Such corrections mainly affect the overall normalization of the cross section but not its energy dependence which is our main interest in this work comparing the linear BFKL and nonlinear BK evolutions.

\section{Numerical results}

\begin{figure*}[tp]
    \centering
    \begin{overpic}[width=1\textwidth]{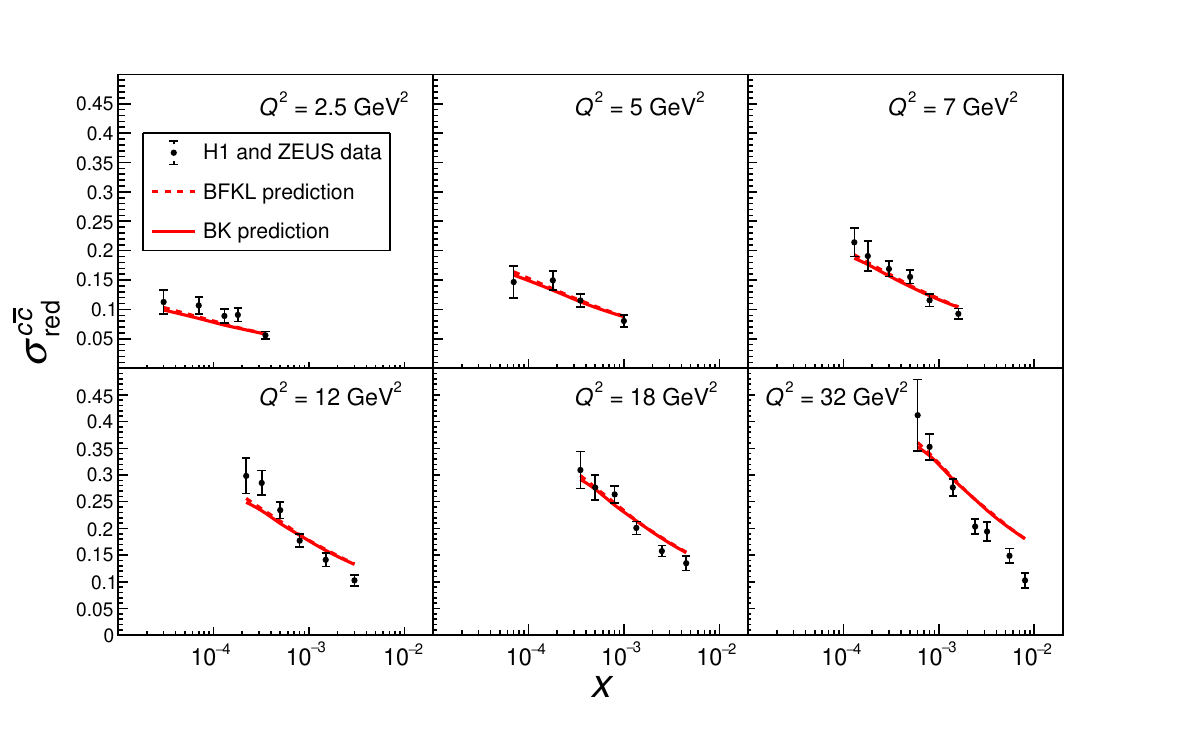}
    \end{overpic}
    \caption{Comparison between reduced charm quark production cross section measurements at HERA and theoretical predictions calculated using Eq.~\eqref{eq:inclusive_quark_production}.}
    \label{fig:inclusive_HERA_comparison}
\end{figure*}

\begin{figure*}[tp]
    \centering
    \begin{overpic}[width=1\textwidth]{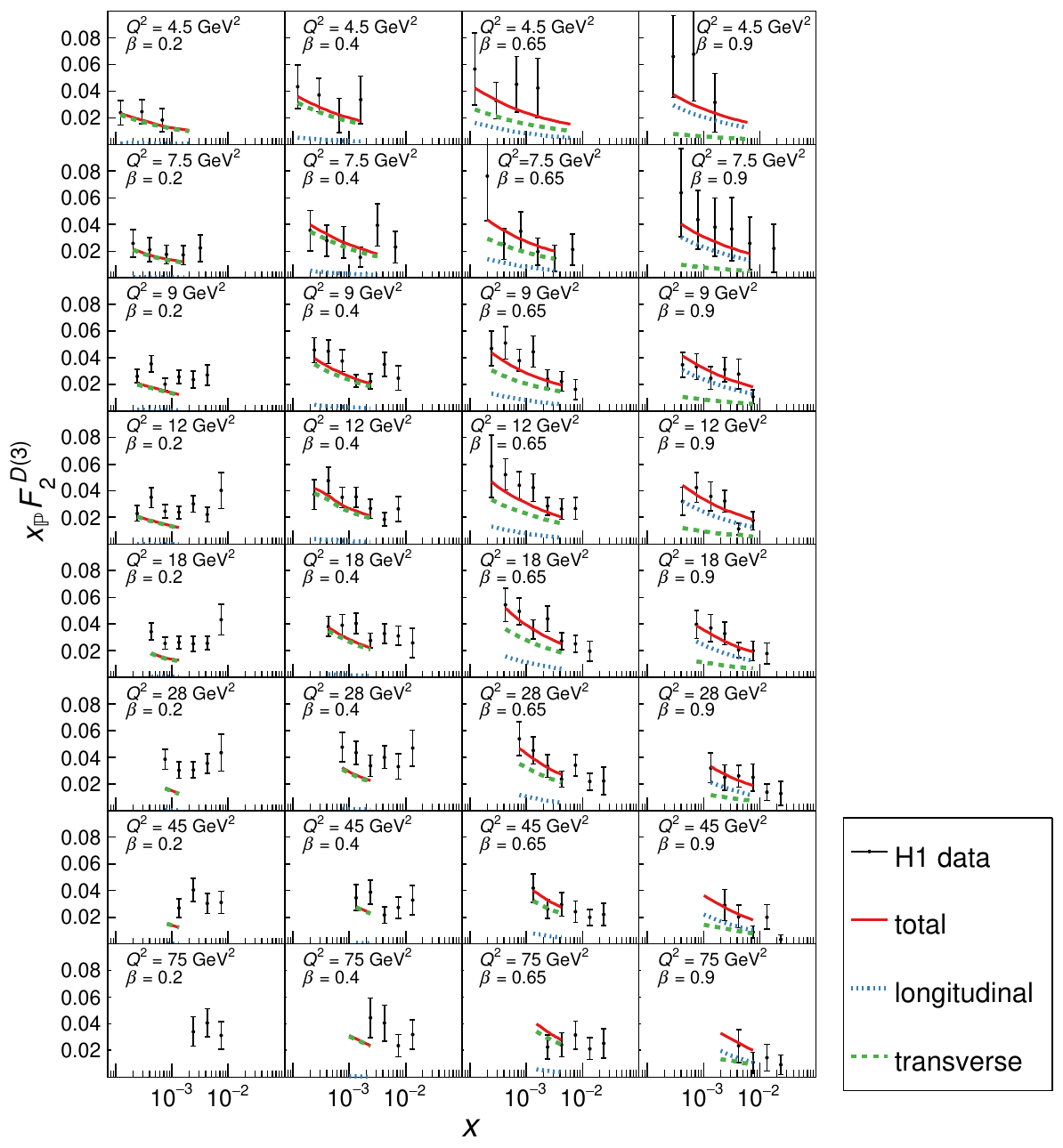}
    \end{overpic}
    \caption{
    Comparison between the $F_2^{D(3)}$ structure function measurements at HERA and the theoretical predictions calculated using Eq.~\eqref{eq:differential_diffractive_quark_production}. Contributions from up, down, strange and charm quarks are included in the prediction. 
    }
    \label{fig:differential_diffractive_HERA_comparison}
\end{figure*}

\begin{figure*}[tp]%
    \centering
    \subfloat[\centering ]{{\includegraphics[width=8cm]{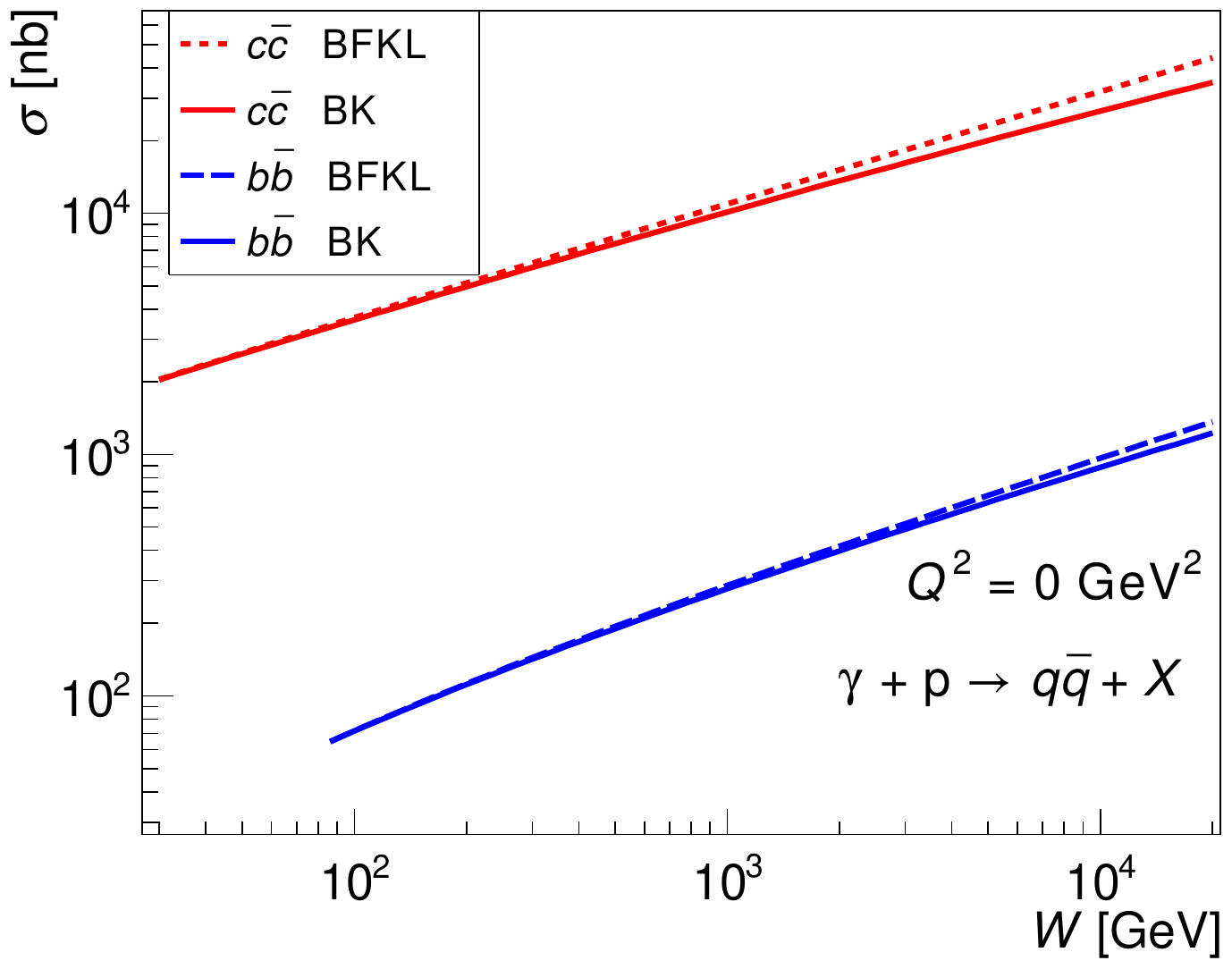} }}%
    \qquad
    \subfloat[\centering ]{{\includegraphics[width=8cm]{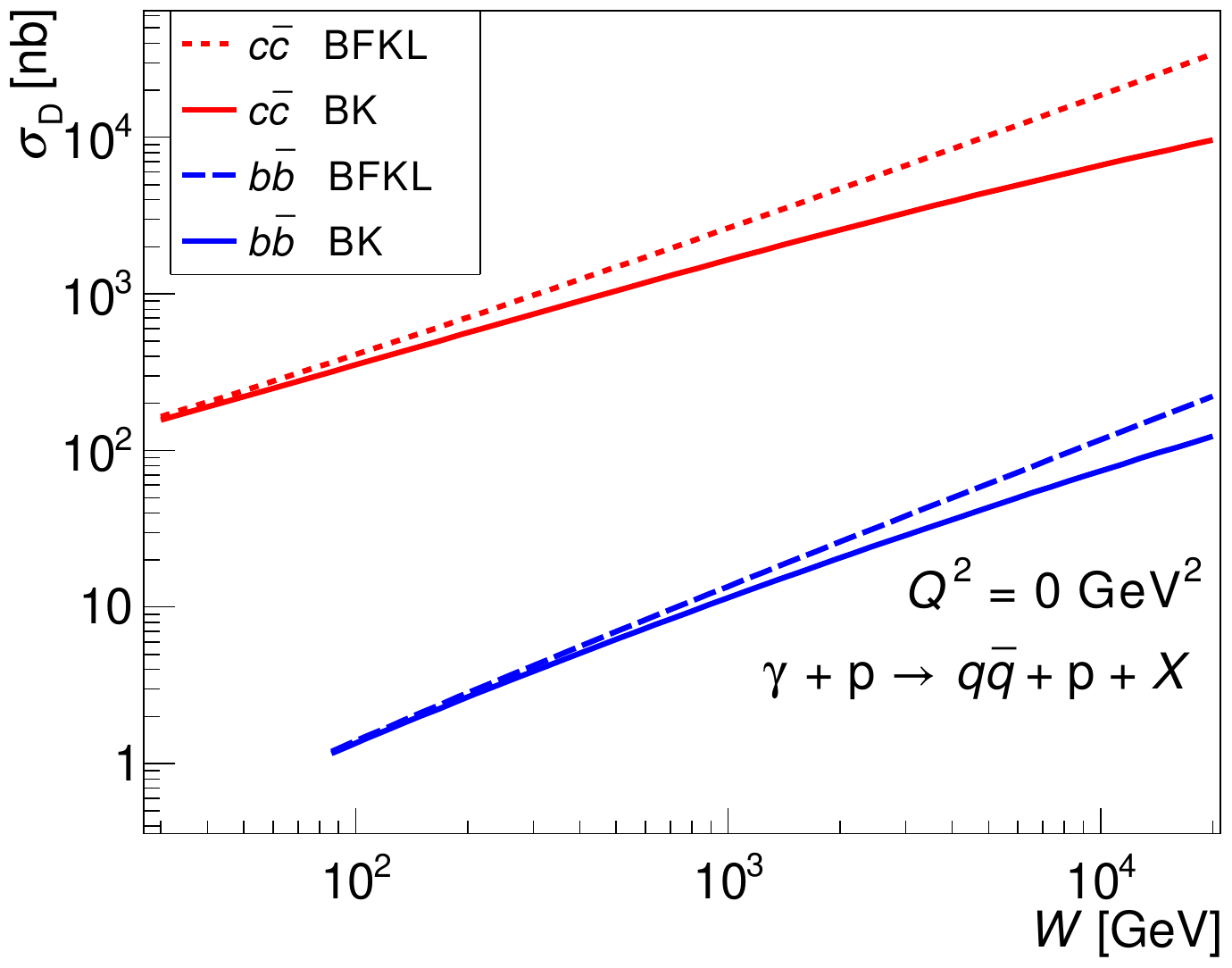} }}%
    \caption{
    Predictions for the heavy quark production cross sections in both inclusive (left) and diffractive (right) scattering off the proton calculated using Eqs.~\eqref{eq:inclusive_quark_production} and~\eqref{eq:diffractive_quark_production}.
    The $b \bar{b}$ and $c \bar{c}$ predictions are shown in blue and red, respectively.}%
    \label{fig:LHC_proton_prediction}%
\end{figure*}

\begin{figure*}[tp]%
    \centering
    \subfloat[\centering ]{{\includegraphics[width=8cm]{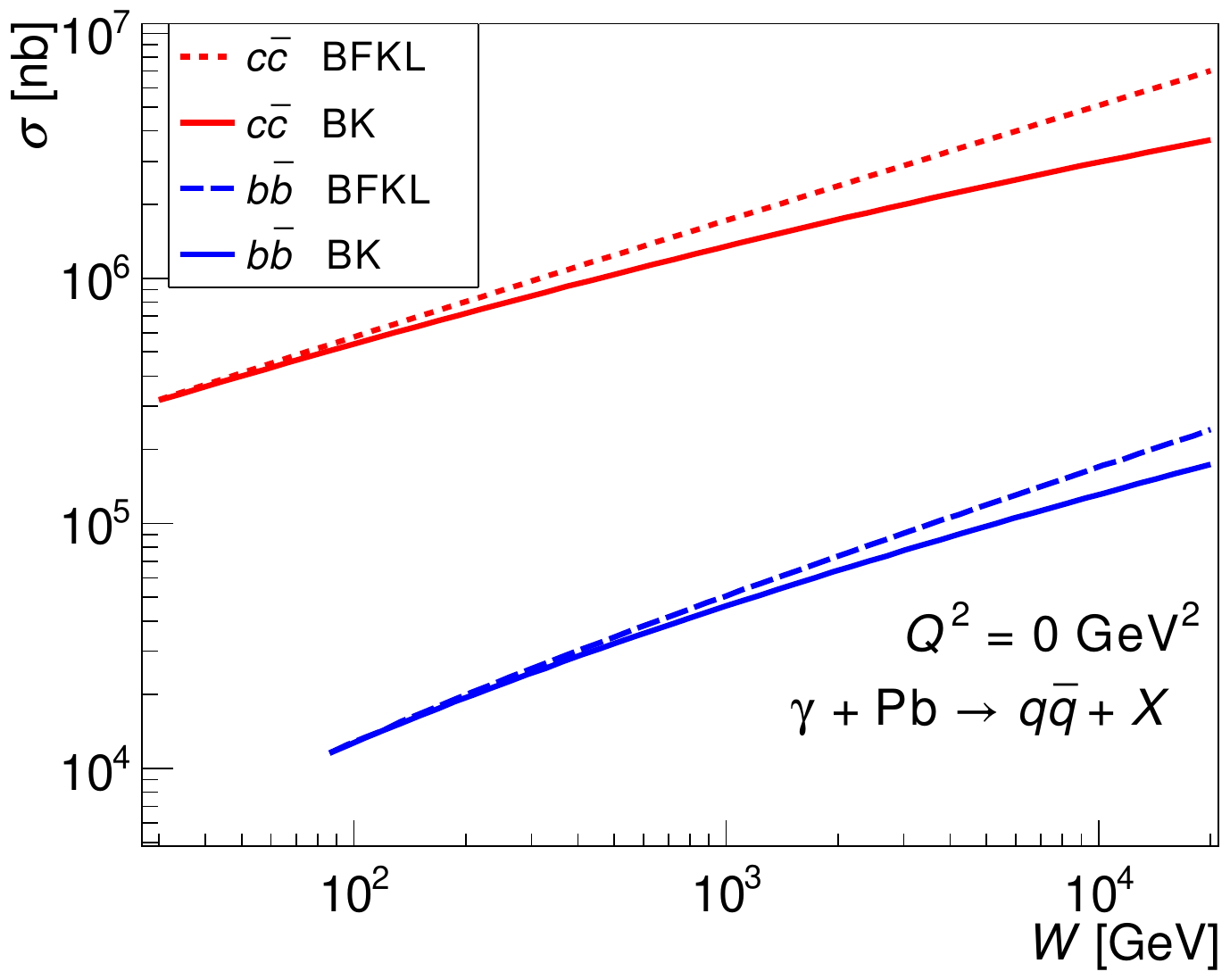} }}%
    \qquad
    \subfloat[\centering ]{{\includegraphics[width=8cm]{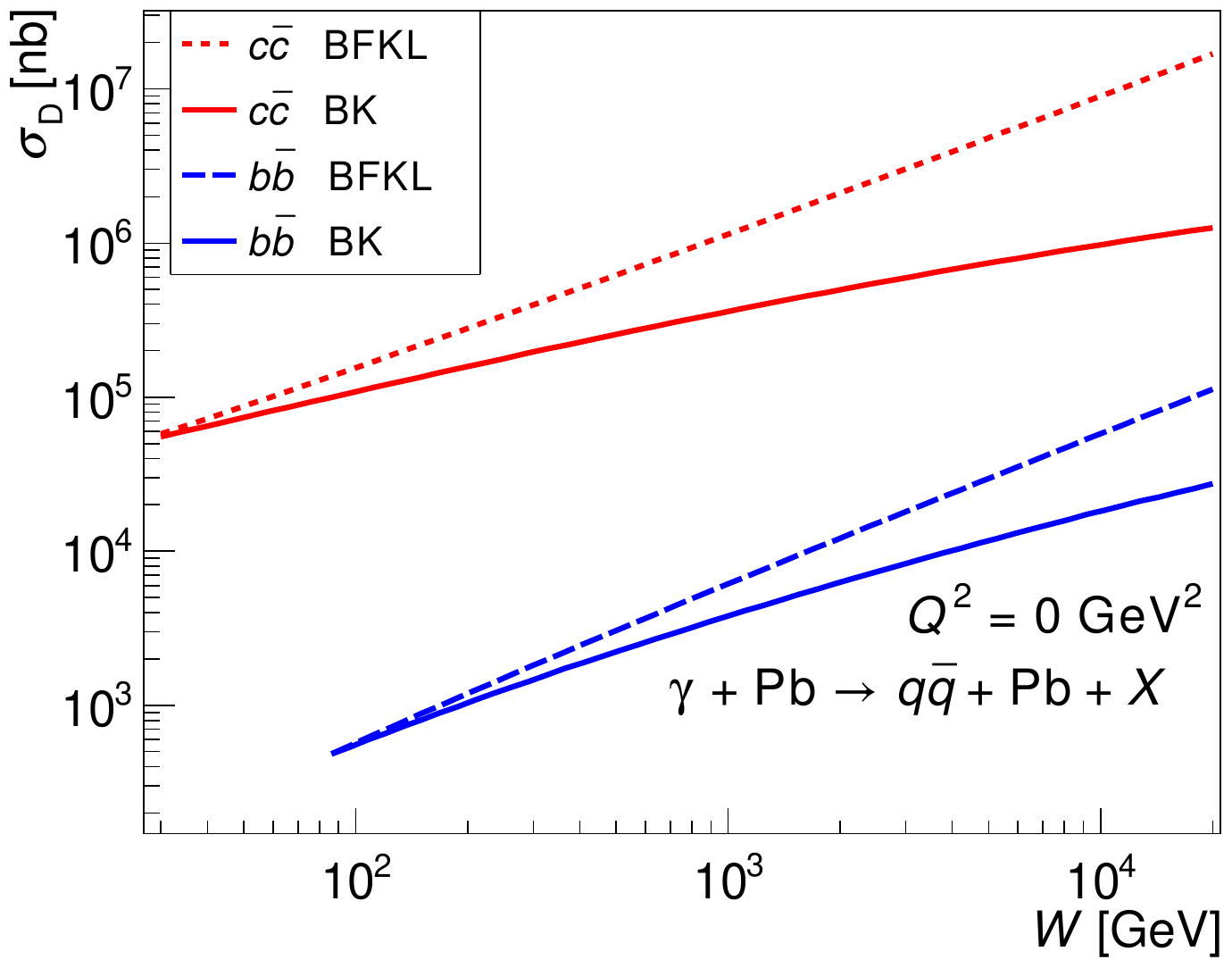} }}%
    \caption{
    Predictions for the heavy quark production cross sections in both inclusive (left) and diffractive (right) scattering off the lead nucleus calculated using Eqs.~\eqref{eq:inclusive_quark_production} and~\eqref{eq:diffractive_quark_production}.
    The $b \bar{b}$ and $c \bar{c}$ predictions are shown in blue and red, respectively.}%
    \label{fig:LHC_lead_prediction}%
\end{figure*}

\begin{figure*}[tp]%
    \centering
    \subfloat[\centering ]{{\includegraphics[width=8cm]{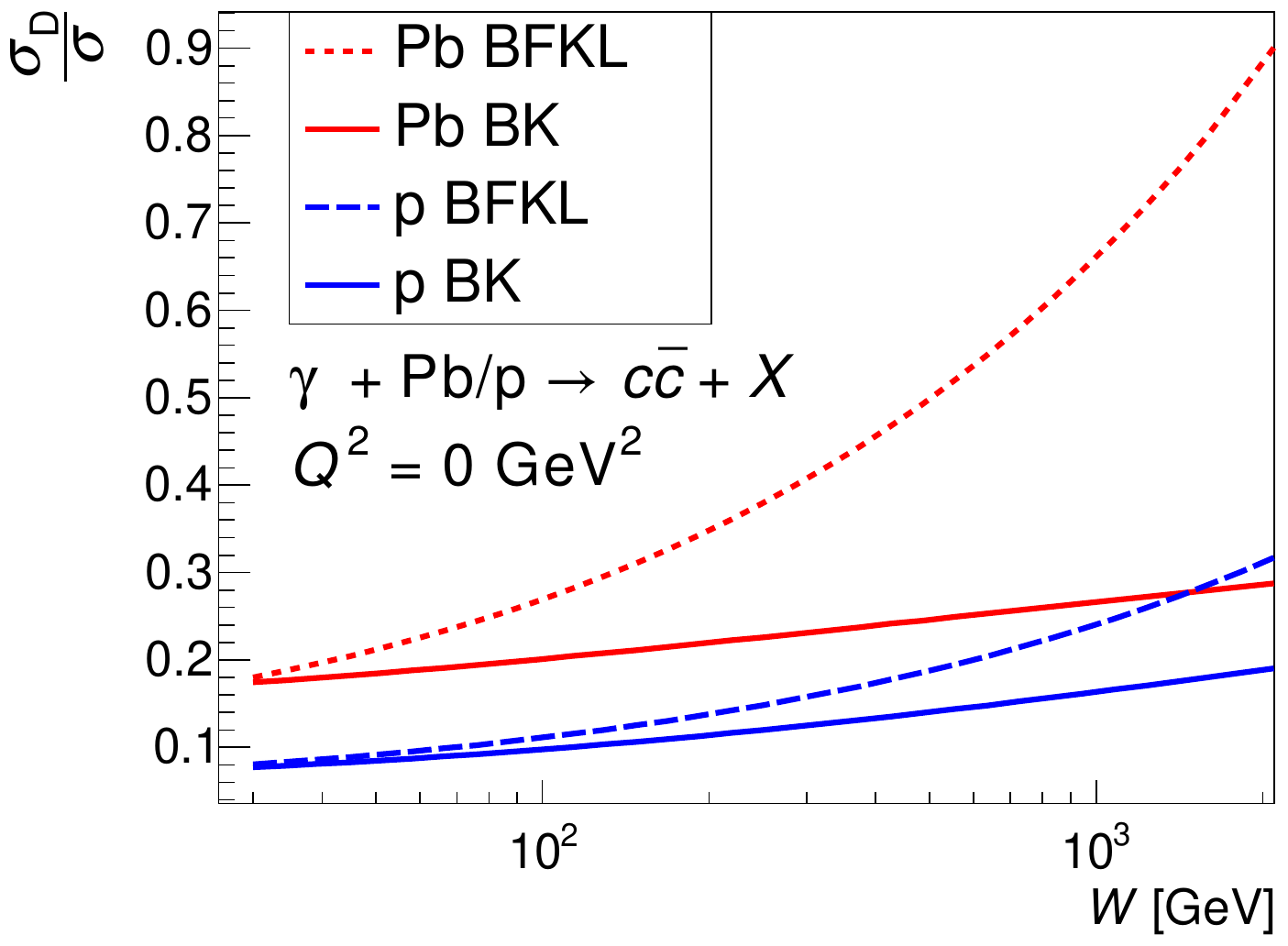} }}%
    \qquad
    \subfloat[\centering ]{{\includegraphics[width=8cm]{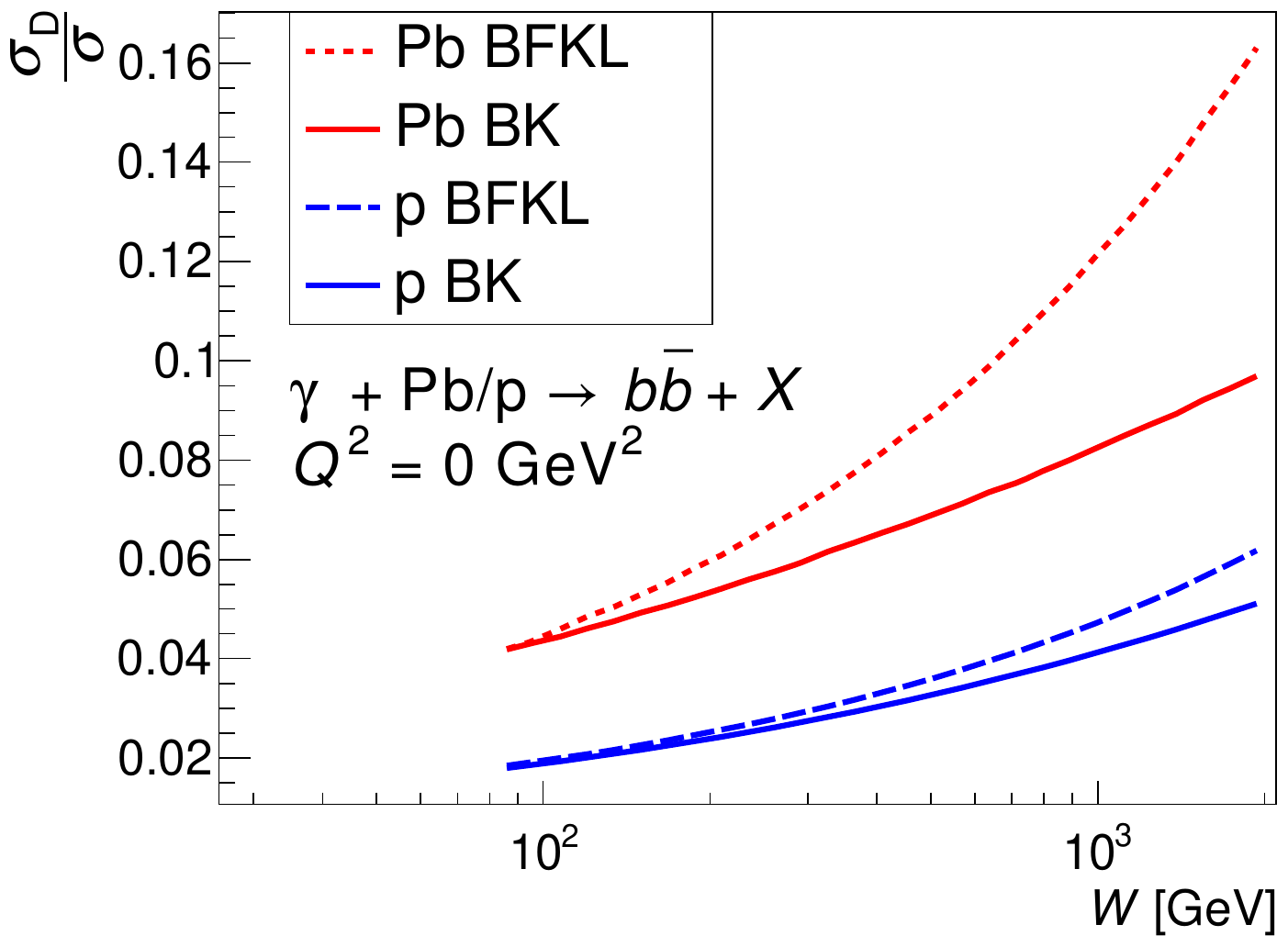} }}%
    \caption{
    Ratios between the diffractive and inclusive heavy quark production cross sections shown in Figs.~\ref{fig:LHC_proton_prediction} and~\ref{fig:LHC_lead_prediction}. Left: $c \bar{c}$ production, right: $b \bar{b}$ production. 
    Predictions for proton and lead targets are shown in blue and red, respectively.}
    \label{fig:diffractive_inclusive_ratios}
\end{figure*}

\label{sec:numerics}

\subsection{Comparison with measurements at HERA}

In Fig.~\ref{fig:inclusive_HERA_comparison}, we present the comparison between our model and the measurement of the $c\bar{c}$ reduced cross section as measured at HERA~\cite{H1:2012xnw}. 
We note that only the parameter $\kappa$ was modified to agree with the overall normalization of the experimental data, and all of the other parameters were obtained by fitting the vector meson data measurements at HERA as in Ref.~\cite{Penttala:2024hvp}. 
We notice a fair agreement between data and our predictions and negligible  differences between the BK and BFKL evolutions. The latter is a consequence of the low gluon density in the target protons, such that $Q^2 \gg Q_s^2$, and we probing the proton in the linear regime.
As expected, the largest differences between data and our model appear at highest $Q^2$ and $x$ where the dipole model is not completely valid.

We also compare our predictions with the measurement of the diffractive structure function $F_2^{D(3)}$~\cite{H1:2006zyl} as shown in Fig.~\ref{fig:differential_diffractive_HERA_comparison}.  
We notice a good agreement between our predictions when $\beta$ is not too small. 
Deviations are observed at medium $\beta$ and $Q^2$ where NLO contributions are known to be important~\cite{Golec-Biernat:1999qor,Kowalski:2008sa,Kaushik:2025roa}.
The differences between the BFKL and BK evolutions are found to be small as expected, showing that the impact of saturation effects at HERA is minimal. 
We have also compared our model with the measurement of the diffractive charm reduced cross section at HERA~\cite{H1:2006zxb}. We get a good agreement between our prediction ($1.08$) and the measurement ($1.50\pm 0.55$) at $Q^2=\SI{35}{GeV}$, $\xpom=0.004$ and $\beta=0.25$, where our formalism is valid.

\subsection{Predictions for charm and beauty production at the LHC}

After comparing our model predictions with the existing data at HERA, we can now compute predictions for $c \bar{c}$ and $b \bar{b}$ production at the LHC in ultra-peripheral $p +\text{Pb}$ and $\text{Pb}+\text{Pb}$ collisions. The $c \bar{c}$ and $b \bar{b}$ cross section predictions are displayed in Figs.~\ref{fig:LHC_proton_prediction} and \ref{fig:LHC_lead_prediction}, respectively, for $\gamma + p$ and $\gamma + \text{Pb}$ interactions, using the linear BFKL and nonlinear BK evolutions. As expected, we see small effects of saturation for $\gamma + p$ interactions due to the small protonic saturation scale (the maximum difference for $W\sim \SI{1}{TeV}$ is of the order of 20\% for the diffractive component) whereas differences up to a factor of 2 are predicted in $\gamma + \text{Pb}$ interactions. 

The ratio of the diffractive to the total $c \bar{c}$ and $b \bar{b}$ cross sections is shown in Fig.~\ref{fig:diffractive_inclusive_ratios}. For $\gamma+ p$ interactions, we predict about 12\% of diffractive events for $W\sim \SI{300}{GeV}$, the HERA center-of-mass energy, which agrees with measurements from the H1 and ZEUS collaborations~\cite{H1:2012pbl,H1:2006zxb}. 
The difference is small between the BK and BFKL evolutions. On the contrary, we predict a large fraction of diffractive events for charm production at the LHC, of the order of 20--25\% for the BK evolution for $W\sim 1$ TeV. 
BFKL evolution leads to about 80\% of diffractive events and even more at higher energies, which shows that the BFKL evolution is not valid at these energies; at high energies, BFKL predictions exceed the maximal ratio of 50\% and break the unitarity of the scattering.
For beauty production, we predict about 7\% of diffractive events at $W\sim$ 1 TeV using the BK evolution, whereas the BFKL predictions yield the fraction 12\%.
Therefore, while the difference between linear and nonlinear evolution is smaller for beauty than for charm due to the larger mass, it should still be measurable at the LHC energies.

\section{Discussion}
\label{sec:discussion}

Measuring $c \bar{c}$ and $b \bar{b}$ production at the LHC in heavy-ion collisions represents an important test of saturation models at the LHC. We predict a difference of about a factor 3.2 between the linear BFKL evolution and the BK evolution that contains saturation for $W \sim \SI{1}{TeV}$ for diffractive production, which is similar to what we have previously predicted in exclusive $J/\psi$ production~\cite{Penttala:2024hvp}. However, the exclusive $J/\psi$ production data can also be described by nuclear and shadowing models without saturation~\cite{Paakkinen:2026dnh,Guzey:2025jfj}.

In order to distinguish between these two different approaches, we propose a measurement of the diffractive production of $c \bar{c}$ and $b \bar{b}$. 
As we have demonstrated, our saturation-based model predicts about 20--25\% of events to be diffractive for $c$-quark production,
whereas nuclear PDF and shadowing models predict a lower fraction similar to the percentage of diffractive events measured at HERA ($\sim 10-15\%$)~\cite{H1:2012pbl,H1:2006zxb}. Such a measurement can thus provide an important test of saturation effects. This can be performed by measuring $D$ and $B$ mesons at the LHC in Pb+Pb collisions, in which the momentum fraction $x$ carried by the charm and beauty quarks can reach very low values~\footnote{It will be possible to perform such a measurement using data taken at the LHC. For instance, The CMS collaboration recorded inclusive Pb Pb interaction data in 2025 without requesting a positive signal in the zero degree calorimeter at trigger level.}.

In addition, measuring UPC vector meson and charm/beauty production at the LHC and the EIC for different nuclear targets will be essential for probing saturation models. The reached kinematical domain is complementary between the LHC and the EIC and will produce measurements with unprecedented precision over a wide domain in energy. The feasibility of running both machines with different ions (He, O, Ar, Ne, Pb, etc...) makes it possible to study the onset of gluon saturation, which will appear at lower momentum scales for heavier ions. 
Comparisons to predictions from collinear factorization, utilizing nuclear PDFs and nuclear shadowing, will provide additional insight into the differences between the linear and nonlinear regimes of QCD.

\section*{Acknowledgments}

J.P is supported by the National Science Foundation under grant No. PHY-2515057, and by the U.S. Department of Energy, Office of Science, Office of Nuclear Physics, within the framework of the Saturated Glue (SURGE) Topical Theory Collaboration.

\bibliographystyle{JHEP-2modlong.bst}
\bibliography{refs}

% \appendix

\end{document}

%% file: packages.tex
\usepackage[utf8]{inputenc}
\usepackage{mathrsfs}
\usepackage{amssymb}	
\usepackage{amsmath}
\usepackage{graphicx}
\usepackage{subdepth}
\usepackage{physics}
\usepackage{xcolor}

\usepackage{subcaption}
\usepackage{siunitx}

\definecolor{lcolor}{rgb}{0.5,0,0}
\definecolor{citcolor}{rgb}{0,0.3,0.0}

\usepackage{overpic}

\usepackage[breaklinks,colorlinks,urlcolor=blue,citecolor=citcolor,linkcolor=lcolor]{hyperref}

\usepackage{graphicx}

%% file: defs.tex
\newcommand{\xt}{{\mathbf{x}}}
\newcommand{\bt}{{\mathbf{b}}}
\newcommand{\rt}{{\mathbf{r}}}
\newcommand{\yt}{{\mathbf{y}}}

\newcommand{\zt}{{\mathbf{z}}}

\newcommand{\nc}{N_\text{c}}

\newcommand{\lambdaQCD}{\Lambda_\text{QCD}}

\newcommand{\ov}{\overline}

\newcommand{\xpom}{x_\mathbb{P}}

\newcommand{\as}{\alpha_\text{s}}
\newcommand{\aem}{\alpha_\text{em}}